\documentclass[twocolumn,secnumarabic,amssymb, nobibnotes, aps, prd]{revtex4-1}

\usepackage{amsmath}
\usepackage{amssymb}
\usepackage{graphicx}
\usepackage[cmyk]{xcolor}
\usepackage{float}
\usepackage{epstopdf}
\usepackage{subcaption}
\usepackage{fullpage}
\usepackage[section]{placeins}
\usepackage{soul}
\usepackage{color}
\usepackage{multirow}

\begin{document}

\title{Dynamic wetting effects in finite mobility ratio Hele-Shaw flow}%

\author{S.J. Jackson}%
\author{D. Stevens}%
\author{D. Giddings}%
\author{H. Power}%
\email[henry.power@nottingham.ac.uk]{}
\affiliation{Faculty of Engineering, Division of Energy and Sustainability, University of Nottingham, UK}
\date{July 21, 2015}

\begin{abstract}

In this paper we study the effects of dynamic wetting on the immiscible displacement of a high viscosity fluid subject to the radial injection of a less viscous fluid in a Hele-Shaw cell. The displaced fluid can leave behind a trailing film that coats the cell walls, dynamically affecting the pressure drop at the fluid interface. By considering the non-linear pressure drop in a boundary element formulation, we construct a Picard scheme to iteratively predict the interfacial velocity and subsequent displacement in finite mobility ratio flow regimes. 
Dynamic wetting delays the onset of finger bifurcation in the late stages of interfacial growth, and at high local capillary numbers can alter the fundamental mode of bifurcation, producing vastly different finger morphologies. In low mobility ratio regimes, we see that finger interaction is reduced and characteristic finger breaking mechanisms are delayed but never fully inhibited. In high mobility ratio regimes, finger shielding is reduced when dynamic wetting is present. Finger bifurcation is delayed which allows the primary fingers to advance further into the domain before secondary fingers are generated, reducing the level of competition.  
%
%
%
\end{abstract}

\maketitle

\section{Introduction}

During the displacement of a high viscosity fluid by a low viscosity fluid, perturbations along the interface greater than a certain wavelength can grow into long fingers, creating highly ramified structures in a process known as viscous fingering. Since Saffman and Taylor in 1958 \cite{saffman1958}, there has been extensive research on viscous fingering occurring in Hele-Shaw cells, where the fluid flows between two thinly separated plates. The mobility of a fluid within a Hele-Shaw cell is defined by the cell separation and the viscosity, giving rise to an intrinsic permeability, analogous to that in porous media flows. The study of viscous fingering in Hele-Shaw cells is often used to provide insight into the more complex problem in porous media, whereby effects such as pore geometry and wetting conditions play a significant role in the interface evolution. 

By assuming that the fluid properties vary sharply over the fluid-fluid interface, the displacement can be considered completely immiscible. One such flow, and the motivation behind the current work is the injection and storage of supercritical $CO_2$ in deep subsurface aquifers ($CO_2$ sequestration). When $CO_2$ is injected and spreads into a deep saline geological formation, it will replace the indigenous fluid (brine), thus creating a two-phase flow system. At the depth of the target formation considered as appropriate for injection, the injected $CO_2$ is in a supercritical condition with a temperature of the order of 50$^{\circ}$C, pressure ranging from 10 to 20 MPa and having a density between 0.4 to 0.8 times that of the surrounding brine. The mobility ratio between the $CO_2$ and brine is of order 10. Under these conditions the two fluids are immiscible, with the total miscibility condition between $CO_2$ and water as 374$^{\circ}$C and 22 MPa. Due to the difference in density between the injected $CO_2$ and surrounding brine, the injected $CO_2$ has a tendency to move upwards as a result of buoyancy forces.

For the site to be suitable for geological storage, the presence of a large enough formation with high permeability at suitable depth is required. To avoid the loss of injected $CO_2$, the target formation needs to be overlain by a layer of very low permeability (cap rock), thus creating both a permeability barrier and a capillary barrier to the upwards migrating $CO_2$. As the rising $CO_2$ plume reaches the cap rock a layer of the injected $CO_2$ spreads under the cap rock at some distance from the injection well, of almost constant thickness. The understanding of the evolution of the lateral spread below the cap rock of this moving layer of $CO_2$ displacing the surrounding brine is the main objective of this work. With this idea we used the analogy between Hele-Shaw and porous media flows.

During immiscible displacement in porous media, the flow regime depends greatly on the capillary number at the interface between the fluids, describing the ratio of viscous driving forces to capillary forces \cite{lovoll2005}. The capillary number, $Ca$ is defined as $\mu U_n / \gamma$, where $\mu$ is the dynamic viscosity of the displaced fluid, $U_n$ is the normal interfacial velocity and $\gamma$ is the surface tension.  At capillary numbers above roughly $1 \times 10^{-4}$ with a mobility ratio greater than one, viscous fingering can occur, where long fingers extend into the porous domain from the injection site \cite{lenormand1988}. At capillary numbers lower than $1 \times 10^{-4}$, capillary fingering can occur, where the plume advances in a 'random walk' fashion, and disconnected globules of the displaced fluid can remain trapped in the pores of the porous media. Capillary fingering is typical of late stage interface advancement in $CO_2$ injection sites, such as the Sleipner field, where capillary numbers can be as low as $1 \times 10^{-10}$ after three years of injection \cite{cavanagh2014}. 

As well as the capillary number, the dynamic wetting conditions of the fluids in the porous media play a crucial role in defining the capillary pressure at the interface between the fluids and the resulting interfacial displacement \cite{weitz1987, Dong2010a}. The contact angle of the displaced fluid with respect to the porous media affects the viscous finger width, which is comparable to the pore size if the displaced fluids wets the medium \cite{stokes1986}. The displaced fluid can also leave behind a trailing film of finite thickness that wets the porous medium, resulting in the capillary pressure having a non-linear dependence on the interfacial velocity \cite{weitz1987}. 

To provide qualitative analysis of the $CO_2$ injection process and explore the dynamic wetting effects discussed above, we present a detailed study of the flow in a Hele-Shaw cell, using fluid properties directly analogous to those found in supercritical $CO_2$ injection, and cell geometry that mimics the continuum length scales in a typical subsurface aquifer. We do not pretend to analyse the full $CO_2$ sequestration problem, but instead aim to give indications of the possible plume patterns that can occur during the $CO_2$ injection. We focus on high capillary number regimes that would typically occur close to the injection site, where the flow can exhibit viscous fingering if the mobility ratio is greater than one.  


At the interface between the fluids two boundary conditions must be met; the capillary pressure jump and the continuity of normal interface velocity. Traditionally, the capillary pressure jump is given by Young-Laplace theory, which describes the pressure jump as the sum of the in-plane curvature and the curvature of the fluid meniscus, multiplied by the surface tension. By assuming that the displaced fluid is perfectly wetting, the meniscus curvature is proportional to $2/b$, where $b$ is the plate spacing. While this boundary condition has been used by many authors and proved successful in matching theory with experiment in the limit of low capillary number flows, in the case of higher capillary number flows the condition must be modified to take into account  non-linear wetting terms \cite{park1984}. 

\begin{figure}
           \begin{subfigure}{0.45\textwidth}
           \hspace{-8.4pt}
                \includegraphics[width=\textwidth]{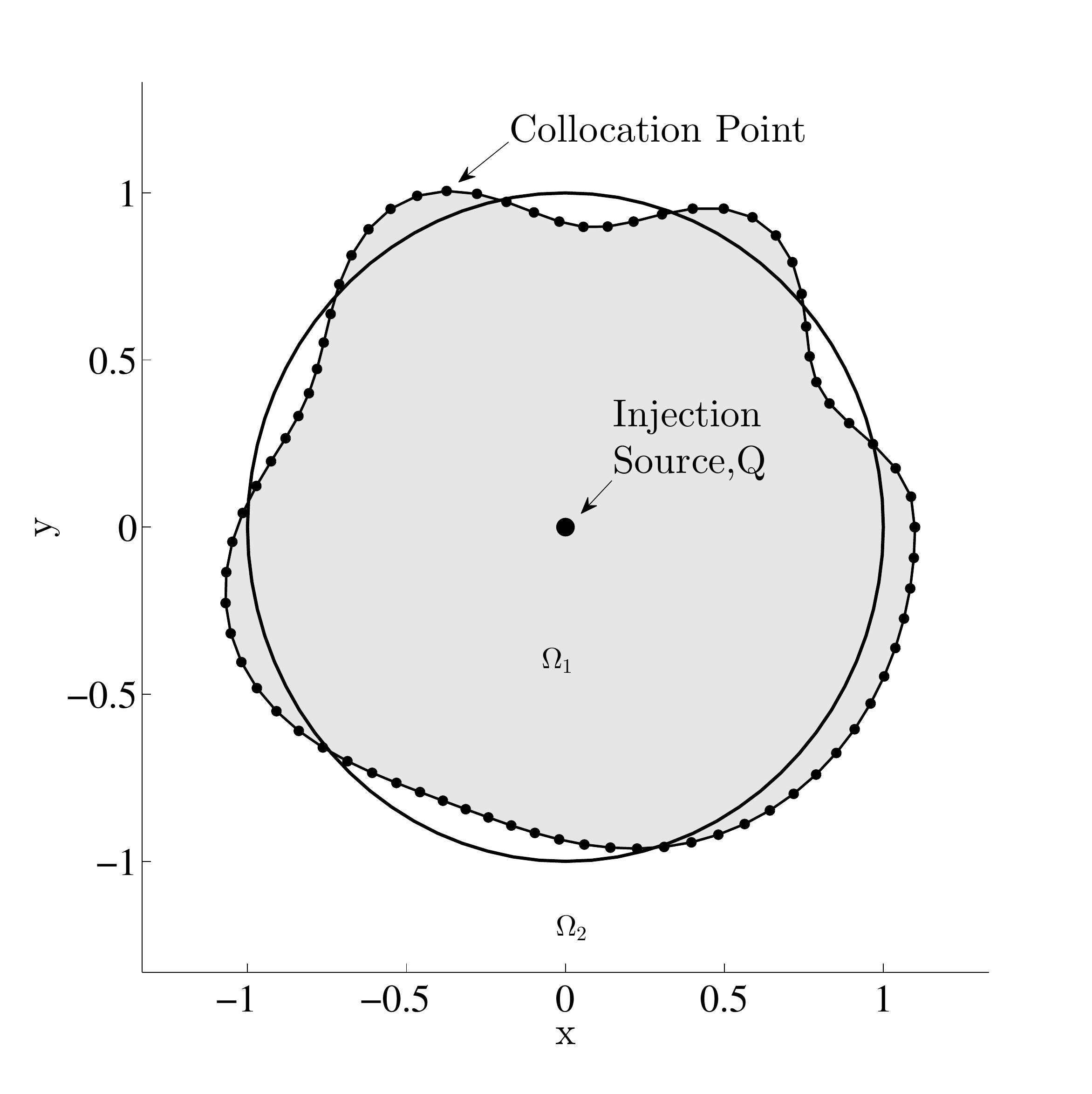}                
                \vspace{-10pt}
                \caption{}
                \label{initial_pertubation}
        \end{subfigure}%
         \\

         \begin{subfigure}{0.4\textwidth}
          \vspace{+10pt}
                \includegraphics[width=\textwidth]{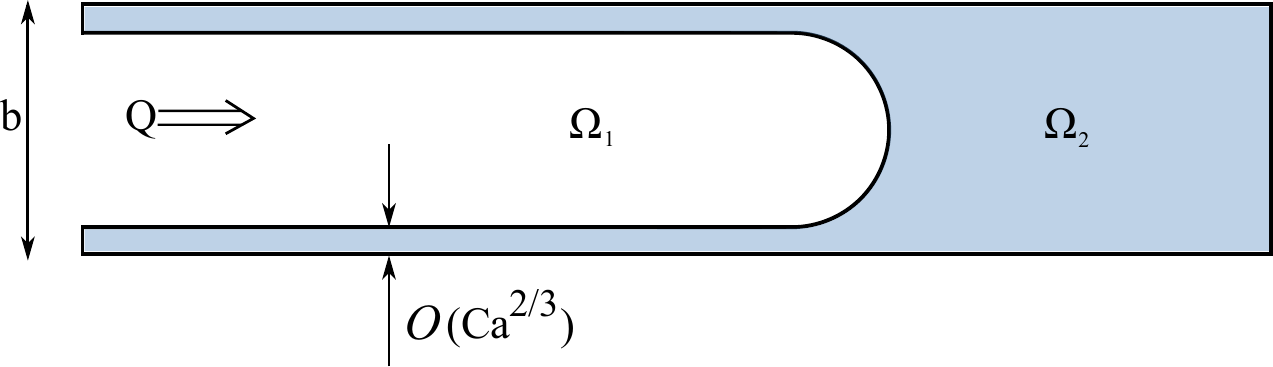}
                \vspace{-15pt}
                \caption{}
                \label{wetting_conds}
        \end{subfigure}%
        \caption{(Color online) (a) Planar view of the 2D radial injection problem, with initial asymmetric boundary. (b) Side view of the Hele-Shaw cell showing trailing film of displaced brine (in blue).}
        \label{initial_setup}
        \vspace{-15pt}
\end{figure}

A non-linear term appears in the pressure jump condition due to the trailing film left behind by the displaced fluid. In the pioneering experimental work by Fairbrother and Stubbs in 1935 for air displacing water in circular capillary tubes, they observed a wetting water film with thickness of the order of $Ca^{1/2}$ in the range $7.5 \times 10^{-5} < Ca < 0.014$ \cite{fairbrother}. Bretherton (1961) studied two phase flow displacement in a two dimensional channel by using a patching analytical approach between the Stokes flow describing the displaced fluid and a lubrication approximation of the thin film left behind the moving interface, as shown in Fig. \ref{wetting_conds}. In the limit of small capillary number, Bretherton found a film thickness of the order of $Ca^{2/3}$ instead of the $Ca^{1/2}$ value found experimentally by Fairbrother and Stubbs \cite{bretherton}.

A formal double asymptotic expansion approach for three dimensional disturbances of a planar flow between two plates was presented by Park and Homsy in 1984, formally proving that in the limit of small capillary number Bretherton’s wetting film thickness of the order of $Ca^{2/3}$ is obtained \cite{park1984}. Besides, the Park and Homsy asymptotic analysis also shows that until order $Ca^{1/3}$, the wetting film thickness does not have a correction term even though the profile of the transition region at the from tip does have a correction at this order. Bretherton infers that a possible explanation for the discrepancy between the theory and the experimental values is the existence of small impurities on the experiments \cite{bretherton}. Recently several works have addressed this issue theoretically and experimentally, showing the possibility of thicker films when small impurities (surfactants) are included in the analysis; see \cite{ratulowski, daripa2011}. As pointed out by Park and Homsy, the effect of the trailing wetting film can be taking into account in the cross average Hele-Shaw cell approximation (Darcy flow) by including in the pressure jump condition across the fluid interface a term linearly proportional to the wetting film thickness, see equation (\ref{eqn7}) \cite{park1984}. 

Since the work of Park and Homsy in 1984, there has been considerable research comparing this modified boundary condition with experimental findings. Early channel flow stability analyses by Schwartz in 1986 and Reinelt in 1987 find that including a non-linear velocity term in the pressure jump improves the agreement between theory and experiment \cite{schwartz1986, reinelt1987}. Schwartz also shows that the linear growth rate of the fingers is reduced when dynamic wetting is included. Later, Maxworthy compares various theoretical boundary conditions for predicting the wavelength of the most unstable wave in a radial Hele-Shaw cell under different capillary numbers and compares these to experimental findings \cite{maxworthy1989}. His results indicate that for high capillary number flows, 3D effects can become significant and could account for the discrepancy between theory and experiment.  

More recently, there has been renewed interest in the stability analysis of the radial injection case \cite{martyushev2011, anjos2013, dias2013}. \cite{martyushev2011} and \cite{dias2013} deal with the linear regime of finger growth, deriving dispersion relations for the growth rate. Without wetting effects, the fastest growing fingers have a wavelength proportional to the square root of the capillary number \cite{patterson1981}. However, \cite{dias2013} find that dynamic wetting effects stabilise perturbations and decrease the growth rate of fingers, similar to the findings of \cite{maxworthy1989}. \cite{anjos2013} takes the analysis further and explores the weakly non-linear regime of finger splitting and competition. They investigate the pattern forming mechanism of the problem, finding that dynamic wetting effects stabilise the fingers and lead to the formation of short stubby fingers in contrast to the classical fan-like patterns \cite{anjos2013}. 

Several papers also exist for the radial injection cases with rotating Hele-Shaw cells \cite{alvarez2006, carillo1999} and lifting Hele-Shaw cells \cite{anjos2014}, showing similar stability analyses and wavelength selection for the linear and weakly non-linear growth regimes. To the authors' knowledge, their does not exist any analysis on the highly non-linear regime occurring after finger bifurcation in radial injection whereby finger competition is prominent and mechanisms such as finger break-off can occur in low mobility ratio flows \cite{jackson2015}. 

We present a numerical scheme based on the boundary element method (BEM) presented in \cite{jackson2015}, to study the morphological changes that occur due to dynamic wetting in the late stages of intefacial evolution. To include the effects of dynamic wetting, we utilise a Picard iteration scheme that iteratively improves the pressure jump boundary condition and subsequent  interfacial velocity calculation at each time step. This scheme allows the effective evaluation of the non-linear term, and the prediction of the late stage pattern forming mechanisms due to dynamic wetting in low mobility ratio regimes found in $CO_2$ injection. 

We proceed by first presenting the mathematical formulation and Picard iteration scheme, followed by numerical results and discussion for various capillary number and mobility ratio regimes. 

\section{Mathematical Formulation}

In this section we present a modified BEM based on that previously presented in \cite{jackson2015}. We will briefly discuss the formulation, focussing on the new dynamic wetting conditions and Picard iteration scheme, referring the reader back to \cite{jackson2015} where necessary.  We consider a Hele-Shaw cell of infinite radius, in which a viscous fluid is displaced by the radial injection of a less viscous fluid. The lower viscosity invading fluid (such as $CO_2$) occupies region $\Omega_1$ whilst a higher viscosity fluid (such as brine) occupies the external region, $\Omega_2$, shown in Fig. \ref{initial_setup}(a). Utilising the characteristic length, time, velocity and pressure of the problem, the field variables can be represented in non-dimensional form: 
\begin{align}
 \label{non_dim1} \left( x,y,r \right)& = a \left(x',y',r' \right) \\
 \label{non_dim2}  t &= \frac{a^2}{Q} t'  \\
 \label{non_dim3} u_i &= \frac{Q}{a} u'_i    \;\;\; i = 1,2 \\
 \label{non_dim4} \left( P,\phi \right)_i &= \frac{Q}{M_2} \left(P',\phi' \right)_i   \;\;\; i = 1,2
\end{align}
In equalities (\ref{non_dim1}) - (\ref{non_dim4}), apostrophes identify non-dimensional variables. $t, u, P$ and $\phi$ represent time, two-dimensional velocity, depth averaged pressure and perturbation pressure respectively. The parameters $a, Q$ and $M_2$ are the unperturbed bubble radius, the radial injection flux and the mobility of the displaced fluid respectively. The fluid mobility in region $i$ is related to the Hele-Shaw plate separation, b, and the fluid viscosity, $\mu_i$ by: 
\begin{align}
 \label{mob_eqn} M_i = \frac{b^2}{12\mu_i} 
\end{align}
The $b^2/12$ term in the mobility ratio refers to the intrinsic permeability of the Hele-Shaw cell, defined by the plate separation. For the flow between two thinly separated plates in a Hele-Shaw cell, the depth averaged pressure and two dimensional velocity in each fluid region can be expressed through Darcy's law:
\begin{align}
 \label{eqn1} u_1' &= -\beta \nabla P_1'\\ 
 \label{eqn2} u_2' &= - \nabla P_2'\\ 
 \label{eqn3} &\nabla \cdot {u'} = 0 
\end{align}
In equation (\ref{eqn1}), $\beta$ is the ratio of mobilities between the two fluids, $M_1/M_2$. For $\beta$ values greater than 1, the inner fluid is less viscous than the external fluid. For infinite mobility ratio models, $\beta = \infty$. With a constant viscosity in each of the two fluid regions, equations (\ref{eqn1}) - (\ref{eqn3}) can be reduced to Laplace's equation. From this point on, the apostrophe of all dimensionless variables will be dropped for clarity, and every variable will be assumed to be in its non-dimensional form, unless otherwise stated. The pressure field can be represented as a sum of the pressures due to an injection potential source, $Q$, and a perturbation term, $\phi_i$:
\begin{align}
 \label{eqn5.0} P_1(x) = \phi_1 - \frac{1}{2 \pi \beta} \ln (r) \\
 \label{eqn5.1} P_2(x) = \phi_2 - \frac{1}{2 \pi} \ln (r)
\end{align}
In equations (\ref{eqn5.0}) - (\ref{eqn5.1}), $r$ is the non-dimensional radial distance from a collocation point, $x$, to the source point located inside the injected bubble. At a boundary point, $\xi$ on the fluid interface, S, between $\Omega_1$ and $\Omega_2$, there are two matching conditions that must be met by the advancing interface. Firstly, continuity of normal fluxes:
\begin{align}
 \label{eqn6} q = \beta \frac{\partial \phi_1}{\partial n} =  \frac{\partial \phi_2}{\partial n}
\end{align}
Secondly, the pressure jump across the interface due to the Hele-Shaw meniscus ($2a/b$ term), the dynamic wetting layer ($Ca_l$ term) and the in-plane curvature ($k(\xi))$ term:
\begin{align}
 \label{eqn7} P_1 - P_2 &= \frac{1}{Ca_g} \left(\frac{2a}{b}(1 + J_0|Ca_l|^{2/3}) + \frac{\pi}{4}k(\xi)  \right)
\end{align}
The pressure jump in equation (\ref{eqn7}) includes both the Hele-Shaw meniscus and the dynamic wetting film layer, with $J_0$ being a constant derived in \cite{park1984} as 3.8. In the above equations, we have introduced the global and local capillary number, $Ca_g$ and $Ca_l$ respectively. The non-dimensional scaling of the problem has been performed using the global capillary number, which is a modified version of the local capillary number specific to radial Hele-Shaw flow and describes the flow in terms of the initial conditions:
\begin{align}
 \label{ca} Ca_g = \frac{ aQ}{\gamma M_2} \;\;\;\;\;\; Ca_l = \frac{ \mu_2 {U}_n}{\gamma}  
\end{align}
The global capillary number, $Ca_g$ in equation (\ref{ca}) is equivalent to that presented by Homsy for rectilinear flow, with the macroscopic length scale and characteristic velocity scale (Q/a) adjusted for the radial Hele-Shaw set-up \cite{homsy1987}. An $a/b$ scaling term has been used to form the global capillary number which relates the initial unperturbed bubble radius to the Hele-Shaw plate separation. This modifies the capillary number to include the effective permeability of the cell. In this work, $a/b \gg 1$. Without any wetting effects, the global capillary number and mobility ratio uniquely describe the radial Hele-Shaw flow \cite{jackson2015}. However, with the non-linear dynamic wetting term, $J_0|Ca_l|^{2/3}$ given in equation (\ref{eqn7}), the problem now also depends on the local capillary number at the interface of the two fluids.

In the present two dimensional Hele-Shaw flow problem, the injected gas spreads radially through the space with an interface velocity that decreases with time as the fluid interface evolves. It  follows that the local capillary number also decreases as time progress. The defined global capillary number is prescribed by the initial value of the local capillary number along the unperturbed bubble radius:
\begin{align}
 \label{cal} Ca_g = Ca_l(0,a)\left(\frac{a}{b}\right)^2 24 \pi 
\end{align}
Where $Ca_l(0,a)$ represents the local capillary number at time $t = 0$ along the unperturbed bubble radius, a. The $24 \pi$ comes from the definition of the mobility and the velocity from the source injection. 

Due to the controversy between the experimental results and theoretical values, in the present work we use the Park and Homsy (1984) pressure jump condition, since it correspondents to the theoretical value at small capillary number \cite{park1984}. Besides, as previously commented, in the present Hele-Shaw two dimensional problem the local capillary number reduces in magnitude as time progress, reaching the limit of a small value at a short time in the evolution. This is in contrast with the cases of flow in a capillary pipe or channel where the fingertip velocity remains constant with an associated constant capillary number. 

Following from the pressure field representation in equations (\ref{eqn5.0}) and (\ref{eqn5.1}), the perturbed pressures, $\phi_1$ and $\phi_2$ can be expressed in terms of their corresponding Green's formulae at the fluid interface \cite{jawson1977}, using the two dimensional fundamental solution, $\phi^*$. Taking the limiting value of the normal derivatives of these equations and subtracting the results from one another, the following second kind Fredholm integral equation can be formed \cite{power1995}.
\begin{align}
 \label{eqn11} -\frac{1}{2}q(\xi) + \left(\frac{1 - \beta }{\beta + 1}\right) \int_s K(y,\xi)q(y)dS_y = g(\xi)
\end{align}
In forming (\ref{eqn11}) we have used the pressure jump and continuity of perturbation flux matching conditions at the interface; for more detail see \cite{jackson2015}. The kernel $K(y,\xi)$ in equation (\ref{eqn11}) is the normal derivative of the fundamental solution of the Laplace equation. The non-homogeneous boundary term, $g(\xi)$, is given by the following hypersingular integral.
\begin{align}
 \label{eqn14} g(\xi) = \frac{\beta}{2 \pi} \int_s f(y) \frac{\partial^2}{\partial n_{\xi} \partial n_y} \left( \ln \frac{1}{R(\xi,y)}\right) dS_y
\end{align}
Where, 
\begin{align}
 \label{fe} f(y) = \frac{1}{1 + \beta} \left( (P_1 - P_2) - \left(\frac{\beta-1}{2 \pi \beta}\right) \ln(r) \right)
 \end{align}
With $(P_1 - P_2)$ given by equation (\ref{eqn7}). Equation (\ref{eqn11}) can be solved using a convergent series for $q$, as long as $0 \leq \beta < \infty$ \cite{power1995}. Using infinitely many terms results in an analytical Neumann convergent series solution. The series can be simplified by taking $ \lambda = \frac{\left(1 - \beta \right)}{\left(\beta + 1 \right)}$ and using a discrete number of terms, $m$, to truncate the solution of equation (\ref{eqn11}).
\begin{align}
 \label{eqn15} q(\xi) = q_0(\xi) + \lambda q_1(\xi) + \dots + \lambda^m q_m (\xi)
\end{align}
The terms in equation (\ref{eqn15}) can be calculated recursively, via the following formulae:
\begin{align}
 \label{neu_first} q_0(\xi) &= -2g(\xi) \\
 \label{neu_m} q_m(\xi) &= 2 \int_s K(\xi,y) q_{m-1}(y) dS_y \:\:\: for \:\:\: m \neq 0
\end{align}
The surface integrals in the boundary integral equations are solved by discretising the boundary using uniform cubic B-Spline boundary elements, first shown in \cite{cabral1990}. The hypersingular integral is solved using a subtraction technique to remove the unbounded terms, shown in \cite{jackson2015}. The movement of the fluid-fluid interface can then be calculated using a forward Euler time stepping approach, where $\Delta L_n(\xi)$ represents the dimensionless distance moved by a boundary point in a single time-step:
\begin{align}
 \label{eqn16} U_n(\xi) = \frac{\Delta L_n(\xi)}{\Delta t} = q(\xi) + \frac{x_i(\xi) n_i(\xi)}{2 \pi r^2} 
\end{align}
From equations (\ref{eqn7}) - (\ref{eqn16}) it can be seen that we require the normal interface velocity, $U_n$ in order to predict the local capillary number in the capillary pressure boundary condition. As this is not known \textit{a priori}, a guessed value is used initially ($U_n^0 = 0$), so that we can predict the capillary number and proceed to work out the normal interfacial velocity. After calculation of this new velocity, the local capillary number can be updated and the process repeated until the updated interface velocity does not change from one iteration to the next. At each Picard iteration, $k$, the normal interface velocity is updated using:
\begin{align}
 \label{picard} \tilde{U}_n^k = \tilde{U}_n^{k-1} + \lambda_r \left( U_n^k - \tilde{U}_n^{k-1} \right)
\end{align}
Where $U_n^k$ represents the value of $U_n$, calculated using (\ref{eqn16}), at Picard step $k$. The relaxation factor, $\lambda_r$ in equation (\ref{picard}) ensures that the velocity does not vary too dramatically with each Picard iteration in highly non-linear cases, where $Ca_l$ is large. After each Picard iteration, the updated velocity is used to work out the local capillary number for the next Picard step, i.e:
\begin{align}
 \label{ca_l} Ca_l = \frac{ \mu_2 \tilde{U}_n^{k-1}}{\gamma}  
\end{align}
The scheme converges once the $L_2$ relative error norm between successive velocity updates is less than a specified tolerance. In all cases, a tolerance of $1 \times 10^{-3}$ was found to be low enough that any further updates to the velocity field were negligible at each timestep, generally requiring around 10-50 Picard iterations. Relaxation factors in the range of 0.1 - 0.5 were used to ensure convergence in high $Ca_l$ cases. As the surface grows at each timestep according to equation (\ref{eqn16}), the number of boundary elements along the interface are adaptively increased to maintain a target element size. The target element size varies with each case (as higher capillary number flows require a finer dataset), but is typically of the order 0.05. 

The numerical scheme is first order accurate in time, as expected from the forward Euler time stepping scheme. Higher order time integration schemes were considered, such as the midpoint and Runge- Kutta methods, however, due to the pseudo steady-state problem being solved, a relatively low temporal resolution could be used without affecting the solution quality, and higher order schemes were deemed unnecessary. Generally, a timestep size of around $\Delta t = 0.01$ is used for most simulations. For more details on the spatial and temporal convergence properties of the BEM used in this work, see \cite{jackson2015}.

This numerical method forms the basis to investigate the effects of dynamic wetting on viscous fingering morphologies during the non-linear stages of finger bifurcation and competition. By setting $J_0$ to zero in the pressure jump condition, we can remove the dynamic wetting term, and see the finger morphology without its effects. In the next section we investigate numerically the effects of dynamic wetting by varying the local capillary number, the global capillary number and the mobility ratio between the fluids.

\section{Numerical Simulations}
\subsection{Local Capillary number effects}

The trailing film left by the displaced fluid in the Hele-Shaw cell has a thickness proportional to $Ca_l^{2/3}$. To investigate how the local capillary number affects the interfacial displacement after bifurcation, $Ca_l$ can be varied whilst maintaining the same global capillary number and mobility ratio. If the viscosity of both the injected fluid and displaced fluid are increased by the same factor, and the Hele-Shaw cell spacing, $b$, is increased by the square root of this factor, the global capillary number can be kept constant whilst varying the local capillary number. In effect, the wetting layer thickness is increased by using more viscous fluids, but the mobility of those fluids is the same through the cell, so the effect of dynamic wetting can be evaluated independently. Without wetting effects, having the same $Ca_g$ and mobility ratio would produce exactly the same interfacial displacement \cite{jackson2015}.   
\begin{figure*}[!ht]
\hspace{-30pt}
           \begin{subfigure}{0.5\textwidth}
                 \includegraphics[scale = 0.35]{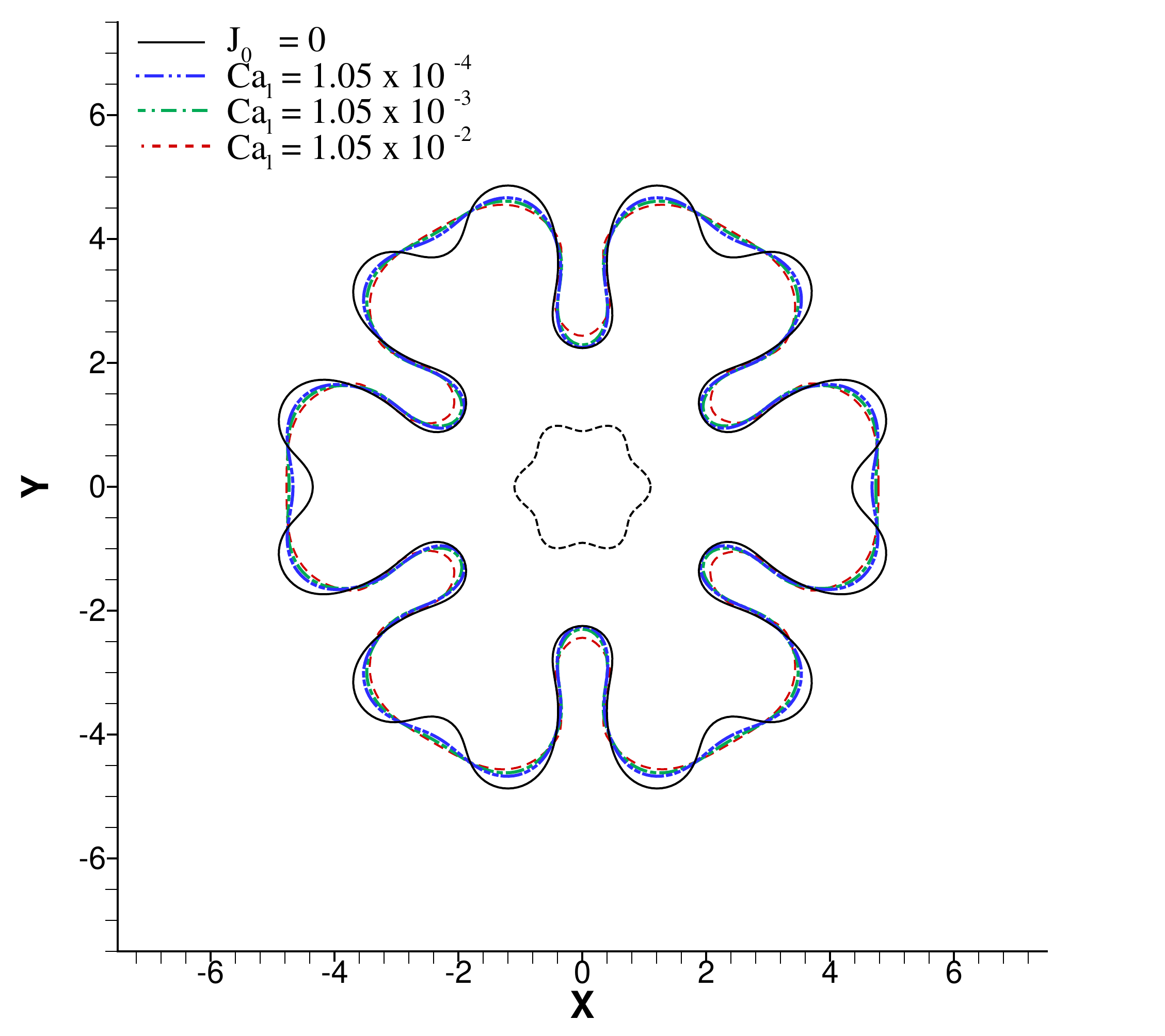}
                \caption{}
                \label{t=55}
        \end{subfigure}%
                 \begin{subfigure}{0.5\textwidth}
                 \includegraphics[scale = 0.35]{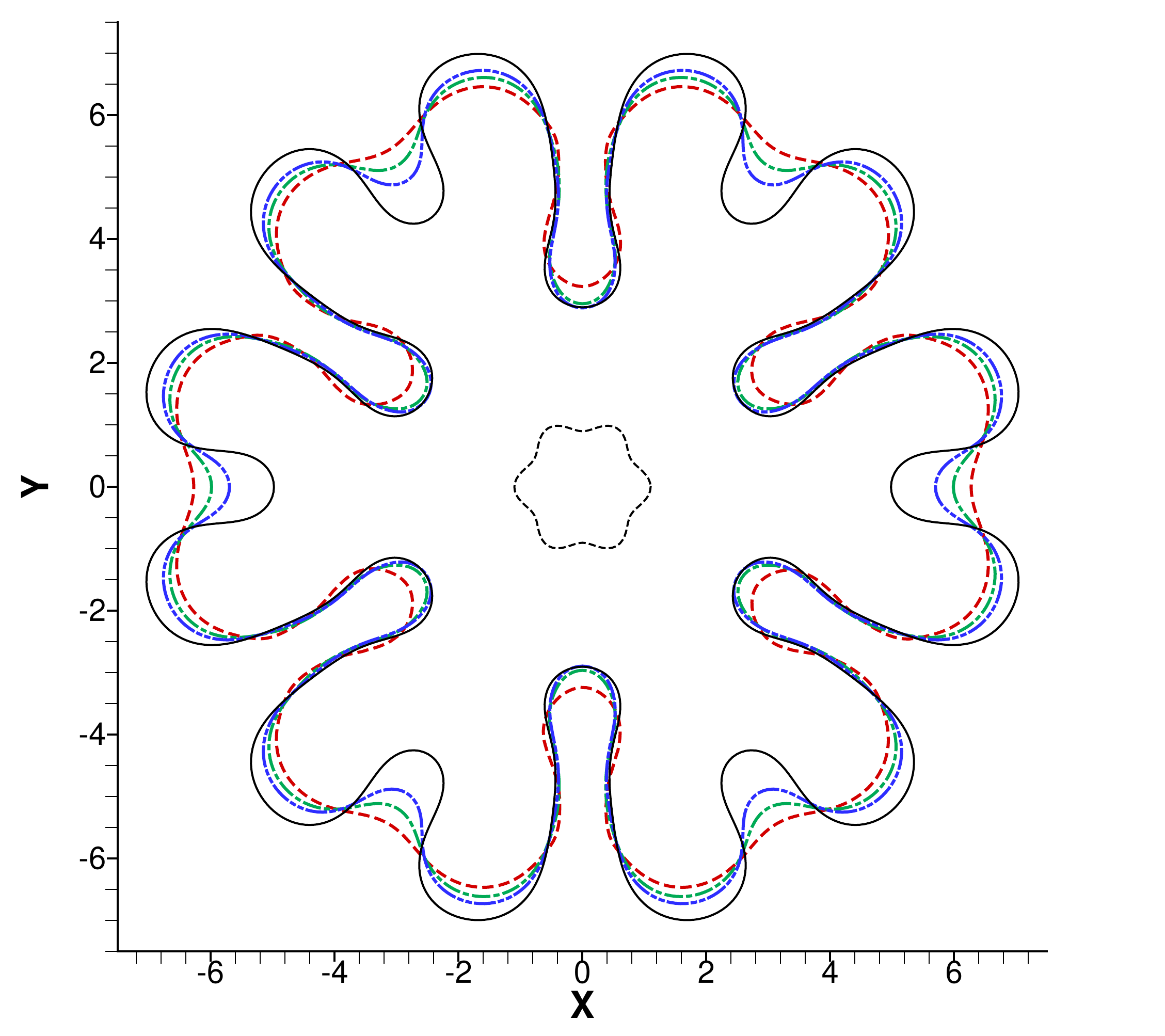}
                \caption{}
                \label{t=110}
        \end{subfigure}%
        \caption{(Color online) Effect of varying the local capillary number on interfacial displacement. (a) t = 55, (b) t = 110. $Ca_g = 1000$, $\beta = 10$.}
        \label{cal_plots}
\end{figure*} 

Several different fluids are injected under the same global capillary number and mobility ratio regime, producing different local capillary numbers. In the results that follow, the local capillary number that is stated in the figures is the maximum found along the interface at t=0, i.e. $|Ca_l(0,r)|_{max}$.

The local capillary number will decay as the radial distance of the interface from the source point increases, but it is convenient to use this initial value for discussion purposes as the local capillary number can vary significantly around the interface at later stages in time. The initial value gives an indication of the level of dynamic wetting that will be present in the regime. By ensuring that $Ca_l$ is above $1 \times 10^{-4}$ initially, we can provide insight into the viscous fingering regime that would occur in a corresponding porous medium subject to the same capillary number flow. 

By altering the viscosity of the fluids, the local capillary number for each simulation is changed. We use brine, low viscosity oil and high viscosity oil as the resident fluids and in each simulation we inject a fluid with a viscosity an order of magnitude lower than the resident fluid, giving a mobility ratio of 10. We focus on low mobility regimes throughout this paper, since our motivation stems from the $CO_2$ injection process which occurs in this low mobility regime.        

When testing capillary number dependence, we use an initial symmetric interface so that dynamic wetting effects can be seen at several locations around the interface. A perturbation term, given by $d \: cos(6\theta)$, is added to the initial unperturbed radius, $a$, to initiate instability at the interface:
\begin{equation}
 \label{sym_bound} r = a+ 0.1a\cos(6\theta)
\end{equation}
In Fig. \ref{cal_plots}, the $J_0 = 0$ interface represents a simulation where the effect of dynamic wetting has been excluded. It can be seen that by including the effect of dynamic wetting, the onset of bifurcation has been delayed. With an increasing local capillary number the thickness of the wetting film increases, which causes the point of bifurcation to be delayed, as a small pressure jump is generated that works to push the bifurcating region outwards (Fig. \ref{cal_plots}(a)). At this low global capillary number, the solution is reasonably stable, generating just two fingers at the first bifurcation due to the relatively large critical length scale of bifurcation. 
                
Dynamic wetting effects also alter the evolution of the finger bases, and can be seen to push the bases further into the domain in Fig. \ref{cal_plots}(b). At low values of $Ca_l$, the bases follow a very similar evolution to that with no dynamic wetting effects, however, above $Ca_l = 1.05 \times 10^{-3}$, the base is pushed significantly away. The effect of dynamic wetting in the $Ca_l = 1.05 \times 10^{-2}$ case is very significant. The finger fronts are displaced less and have a much smaller perturbation in comparison to the other cases, causing the bases of the fingers to be pulled much further forwards into the domain.

The sides of the primary fingers near the bases pinch inwards in the non-dynamic wetting case, which is delayed in the other cases due to a small positive velocity being added to the normal interface velocity. It is worth noting that in our model, dynamic wetting has no historical hysteresis, that is, if the finger moves back into a region it has previously been, the wetting layer behaves exactly the same as if the finger had never been there (hence the absolute value of $Ca_l$ in the pressure jump condition). This assumption has been discussed in \cite{alvarez2006} concluding that thorough experimental work is needed to test the validity of this assumption. 

The results in Fig. \ref{cal_plots} show very similar properties to the analytical results of Dias and Miranda \cite{dias2013}. In their case, several modes of perturbation are analysed, with displacement computed for the early stages of finger growth and the weakly non-linear stage of bifurcation (finger splitting). They use local capillary numbers in the order of 0.01 - 0.1 (initially), and find that the dynamic wetting layer hinders finger growth and can almost completely inhibit bifurcation, creating short stubby fingers. We find a similar results here, but find that in the later stages of finger growth, bifurcation can still occur even with large values of local capillary number. 

\begin{figure}
\includegraphics[scale=0.6]{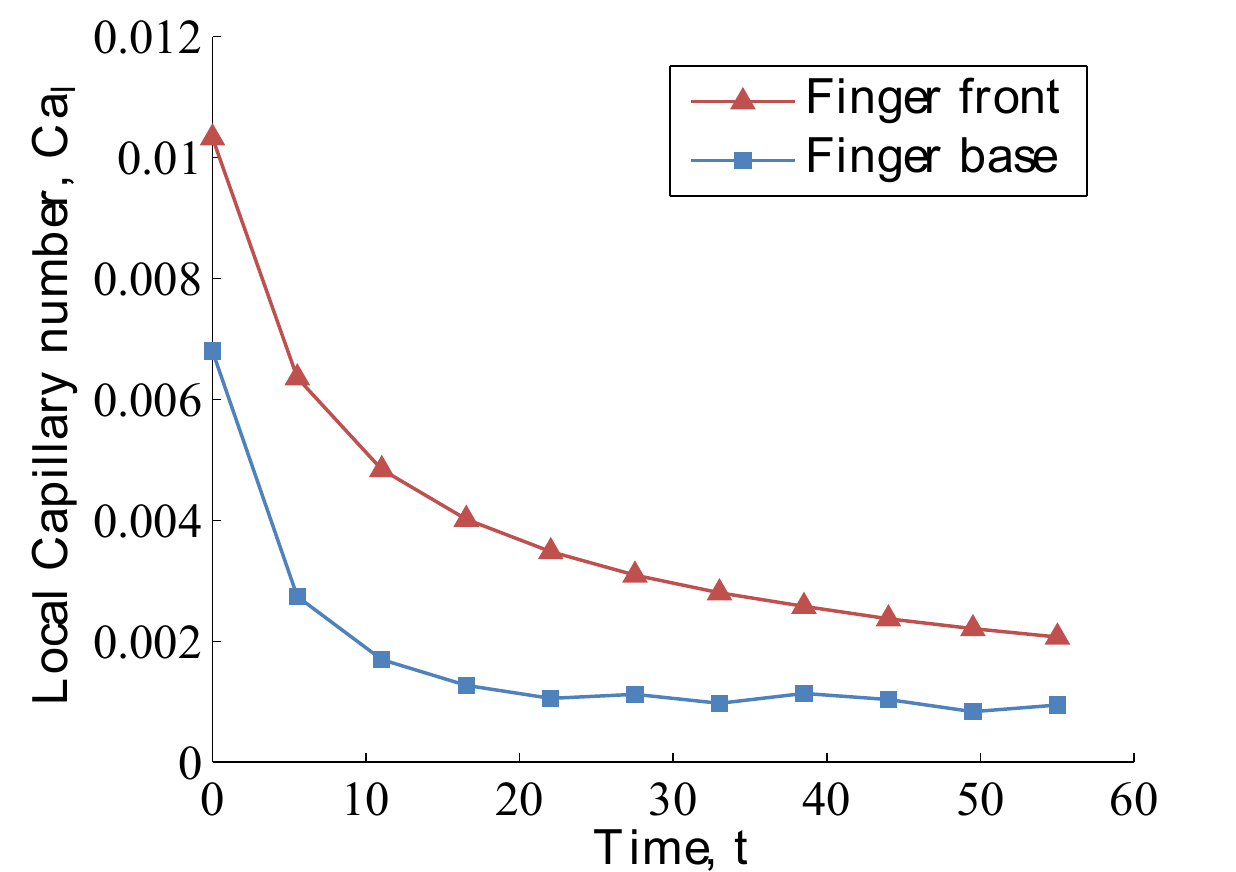}
\caption{(Color online) Local Capillary number variation with time for the dashed line case ($Ca_l$ initial $= 1.05 \times 10^{-2}$) in Fig. \ref{cal_plots}.}
\label{cal_time}
\end{figure}
The local capillary number decreases with time, meaning in later late stages it can be small enough that the usual viscous fingering dynamics occur and bifurcation will proceed. If the bifurcation is delayed significantly, the mode of bifurcation can change and the resulting interface can be significantly different. 

In Fig. \ref{cal_time}, it is reported how the local capillary number changes with time for the highest initial $Ca_l$ case from Fig. \ref{cal_plots}. The local capillary number decreases quickly, and reaches the small capillary number limit of the Park and Homsy correction used in equation (\ref{eqn7}) early in the simulation. Even in the highest initial $Ca_l$ case, the Park and Homsy small capillary number limit is quickly reached, showing the applicability of the correction term for the radial injection cases presented here.

\subsection{Global Capillary number effects}

In this section, we investigate the effect of changing the global capillary number, and how the dynamic wetting layer alters the resulting interfacial evolution. Increasing the global capillary number lowers the critical length scale of bifurcation and generally means that more fingers are generated on the first bifurcation, leading to a much more convoluted interface. With an increase in $Ca_g$, the local capillary number is also increased, so we see the effect of both a more unstable solution globally, and an increased thickness of dynamic wetting layer.

\begin{figure}
          \begin{subfigure}{0.38\textwidth}
                \includegraphics[width=\textwidth]{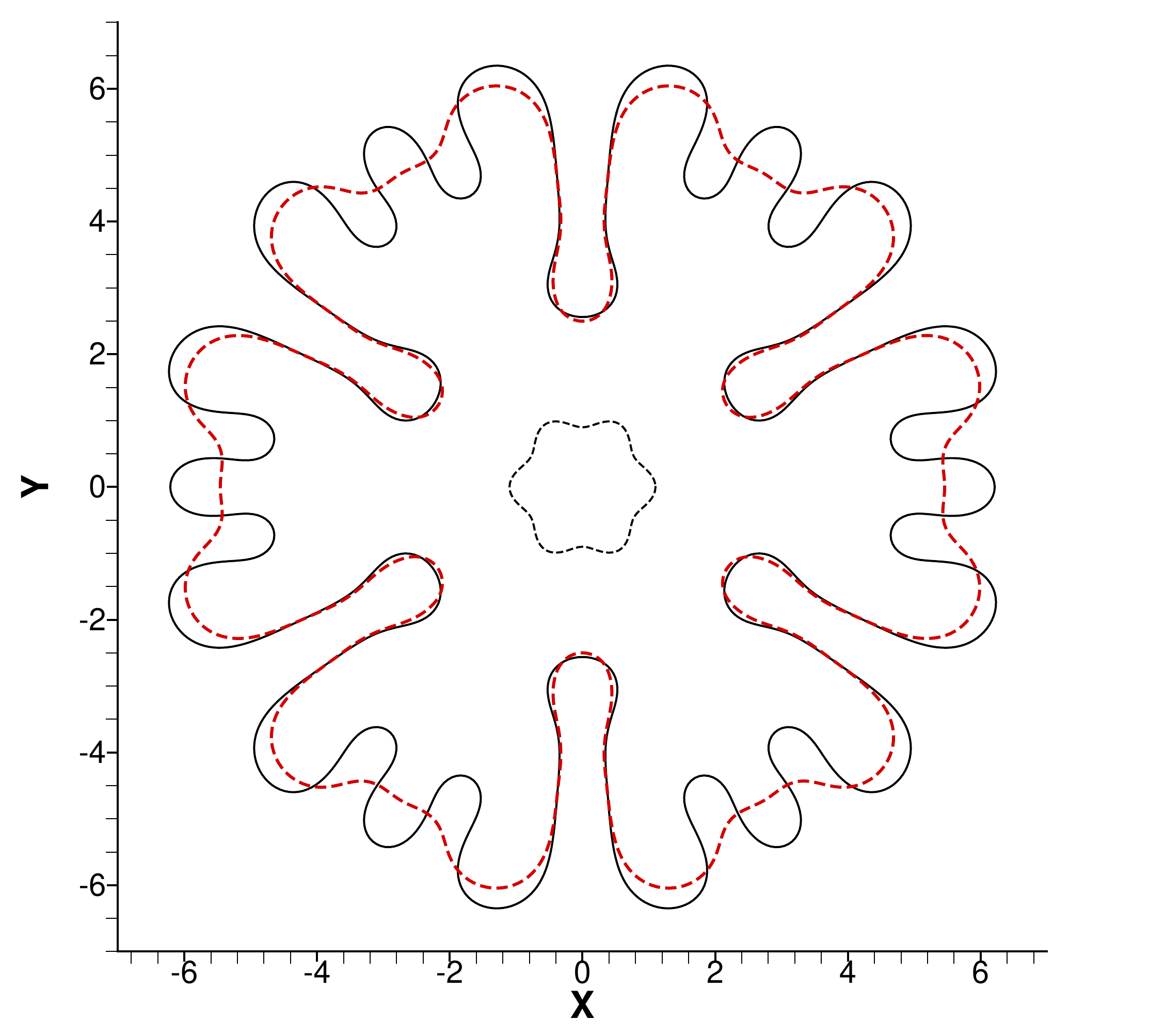}
                \caption{}
                \label{ca_2k}
        \end{subfigure}%

         \begin{subfigure}{0.38\textwidth}
                \includegraphics[width=\textwidth]{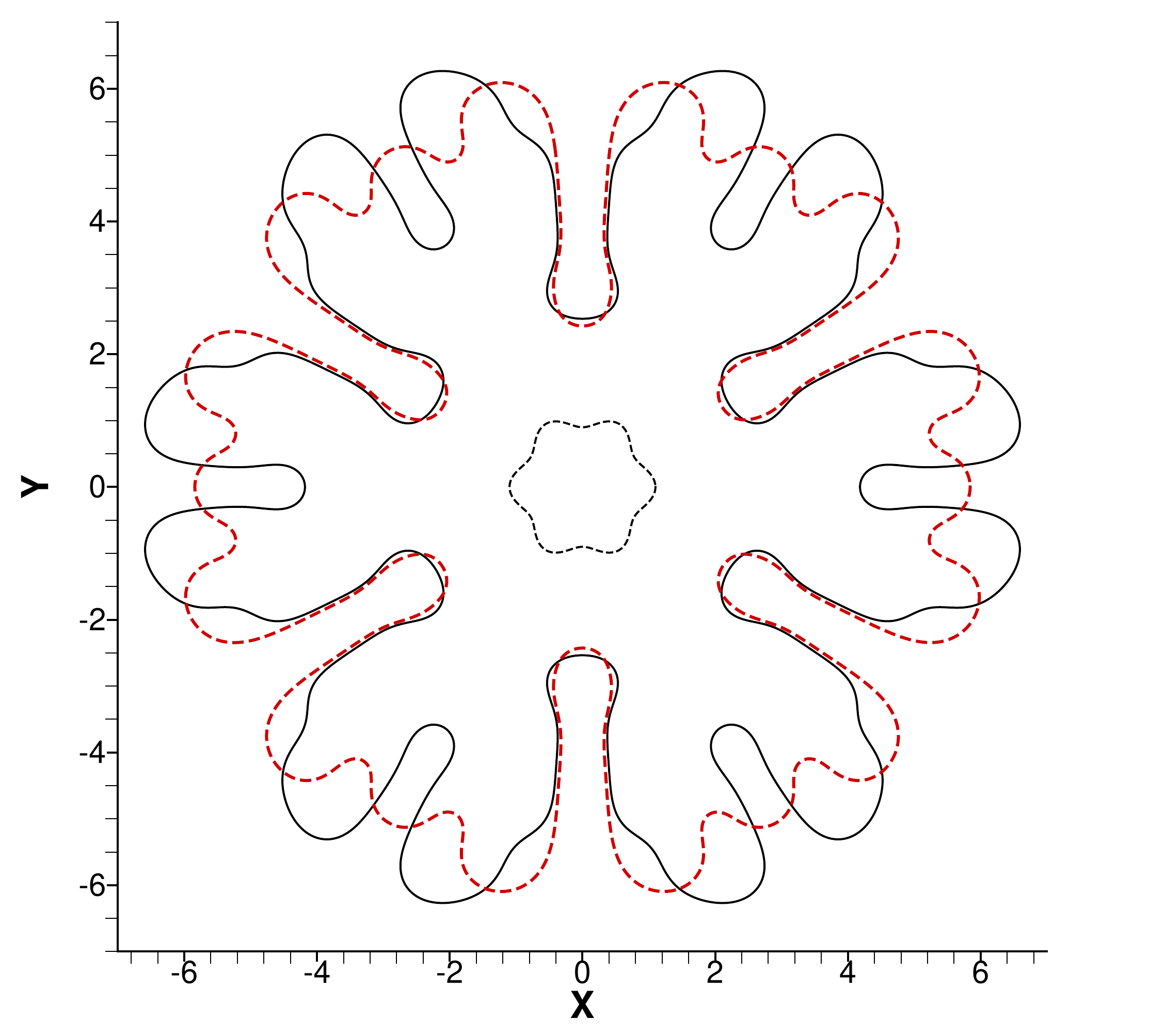}
                \caption{}
                \label{ca_4k}
        \end{subfigure}%
        
                 \begin{subfigure}{0.38\textwidth}
                \includegraphics[width=\textwidth]{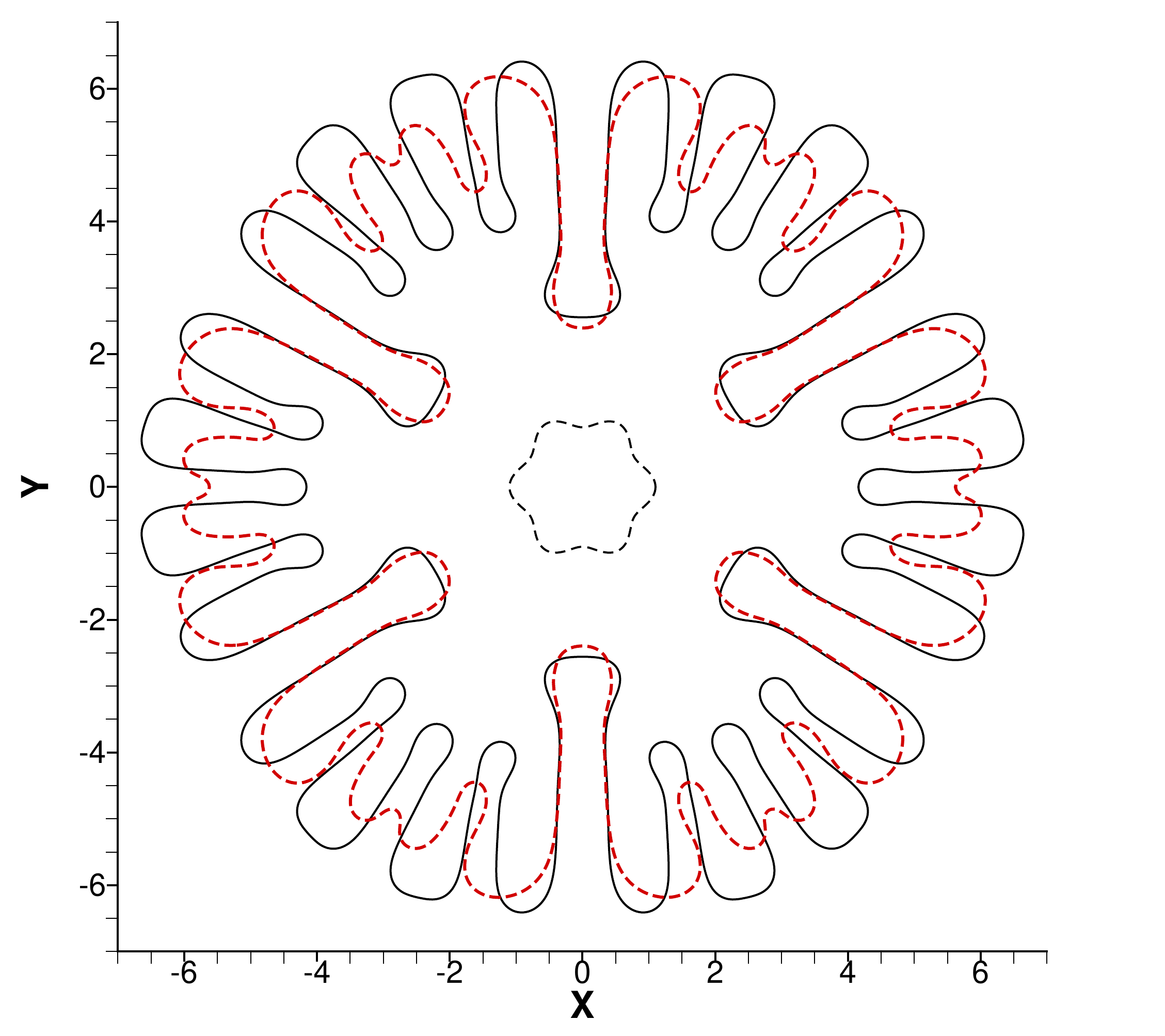}
                \caption{}
                \label{ca_8k}
        \end{subfigure}%
        
        \caption{(Color online) Effect of varying $Ca_l$ and $Ca_g$ on interfacial evolution, t = 88. \\ (a) $Ca_g = 2000$. Initial $Ca_l = 2.19 \times 10^{-4}$ \\ (b) $Ca_g = 4000$. Initial $Ca_l = 4.47 \times 10^{-4}$ \\ (c) $Ca_g = 8000$. Initial $Ca_l = 9.04 \times 10^{-4}$ \\\textbf{\textemdash} $J_0 = 0$, {\color{red} \textbf{-}\textbf{-}\textbf{-}\textbf{-}} $J_0 = 3.8$. $\beta = 10$}
        \label{global_ca}
\end{figure}

The changing global capillary number results are presented in Fig. \ref{global_ca}. We use a mobility ratio of 10 in all cases, with brine as the displaced fluid, and supercritical $CO_2$ as the injected fluid. To increase the global capillary number we increase the injection flux, Q. The local capillary number stated in each figure refers to the initial maximum along the interface calculated at t = 0. In all three examples it can be seen that dynamic wetting delays the onset of bifurcation. The small positive pressure jump induced by the dynamic wetting layer allows the bifurcating point to remain close to the rest of the finger front.

With the relatively high local and global capillary numbers tested in Fig. \ref{global_ca} it can be seen that the bifurcation is delayed, and the fundamental mode changed, when dynamic wetting effects are included.  In the $Ca_g = 2000$ case, the three finger split has been inhibited, and only two primary fingers form from the first bifurcation. The bifurcation has been delayed, and in doing so the fundamental splitting mode has been changed due to the inhibition of the middle finger.

We see similar effects at higher global capillary numbers, although the number of fingers that form on the first bifurcation is not always reduced due to dynamic wetting. The critical length scale of bifurcation increases at higher capillary numbers, meaning it is easier for more fingers to form upon bifurcation. However, as the interface is generally 'smeared' and made more uniform by the inclusion of dynamic wetting, there can be more parts of the interface greater than the critical length scale of bifurcation. This means more fingers can sometimes be formed due to the presence of the dynamic wetting layer (see Fig. \ref{global_ca}(b)). 

When the global capillary number is increased, the overall stability of the solution is reduced. The interface becomes much more ramified and many more boundary elements must be used to accurately capture the displacement. However, the smoothing effect of dynamic wetting also increases with global capillary number, meaning that there is competition between the destabilising effect of increasing the viscous driving force and the stabilising effect of increasing the thickness of the dynamic wetting layer. Through many numerical experiments, it was observed that even at very large local capillary numbers, bifurcation could never be completely inhibited. The system would eventually become unstable due to the viscosity contrast between the fluids and the fact that the wetting layer thickness reduces with time.

\subsection{Late stage interfacial displacement}

Here we consider late stage interfacial evolution, given an asymmetric starting interface in order to more closely represent the different perturbation wavelengths that may occur naturally in a real injection scenario. We use the asymmetric boundary in Fig. \ref{initial_setup}(a), defined by:
\begin{equation}
 \label{asym_bound} r = a + 0.1a \cos \left( 6 \sqrt{\frac{\theta^{3.5}}{2 \pi^{2.5}}} \right)
\end{equation}
We consider the displaced fluid to be brine, with suitable deep aquifer properties (injection depth 1000m, $P_{amb} = 10MPa$, $T = 30 ^{\circ}C$, $\gamma = 0.03 kg.s^{-2}$, $\mu = 7.6 \times 10^{-4}Pa.s$). Supercritical $CO_2$ is injected into the system with a pressure of $20MPa$, and a viscosity ten times lower than that of the resident brine. This gives a global capillary number of 5000, with a initial local capillary number of $5.33 \times 10^{-4}$. The plate separation is $0.0123cm$, giving an intrinsic permeability of $1.26 \times 10^{-5} cm^2$ equivalent to highly fractured porous rock \cite{bear1972}.

\begin{figure}
\centering
           \begin{subfigure}{0.4\textwidth}
                \includegraphics[width=\textwidth]{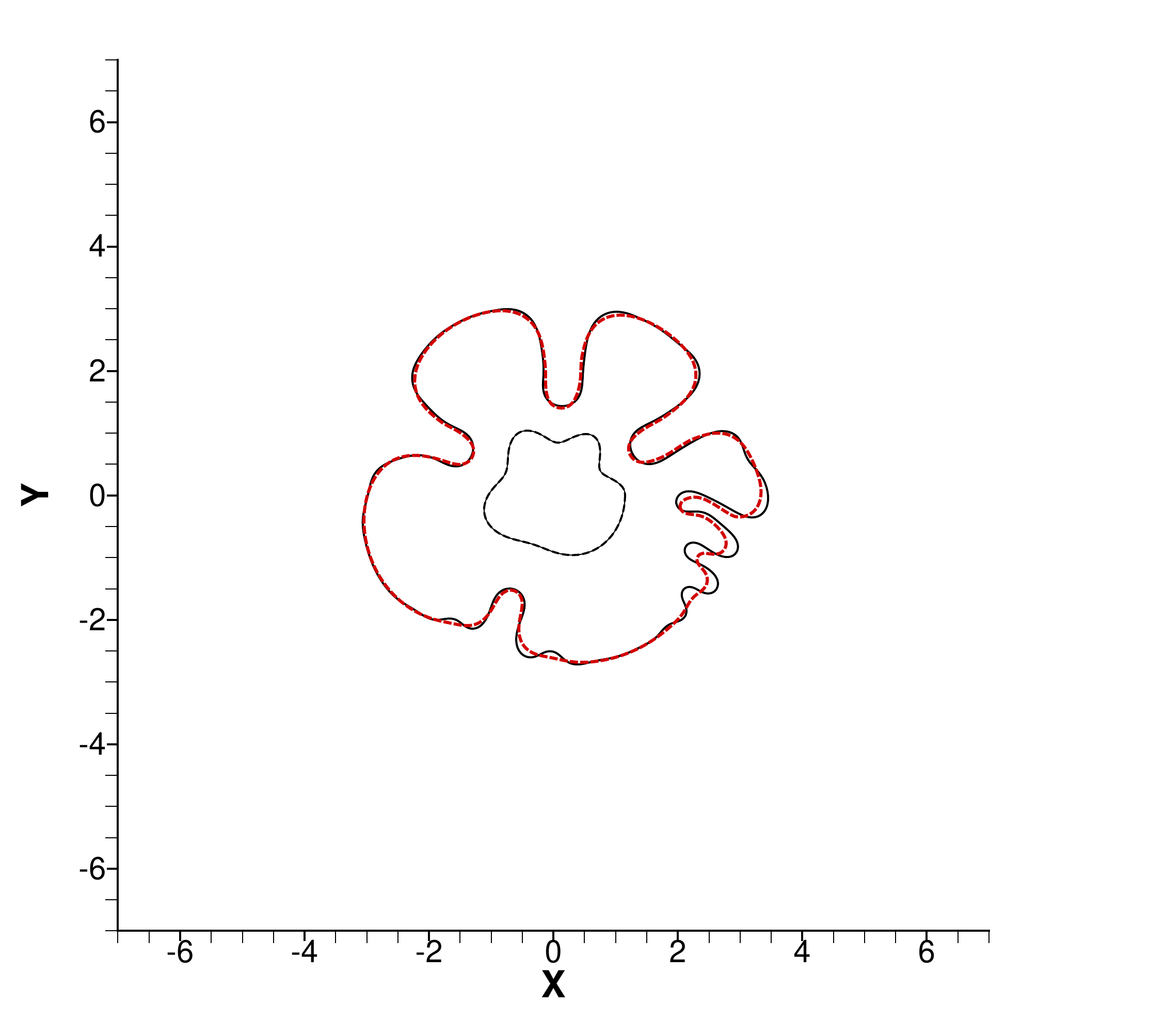}
                \caption{}
        \end{subfigure}%

         \begin{subfigure}{0.4\textwidth}
                \includegraphics[width=\textwidth]{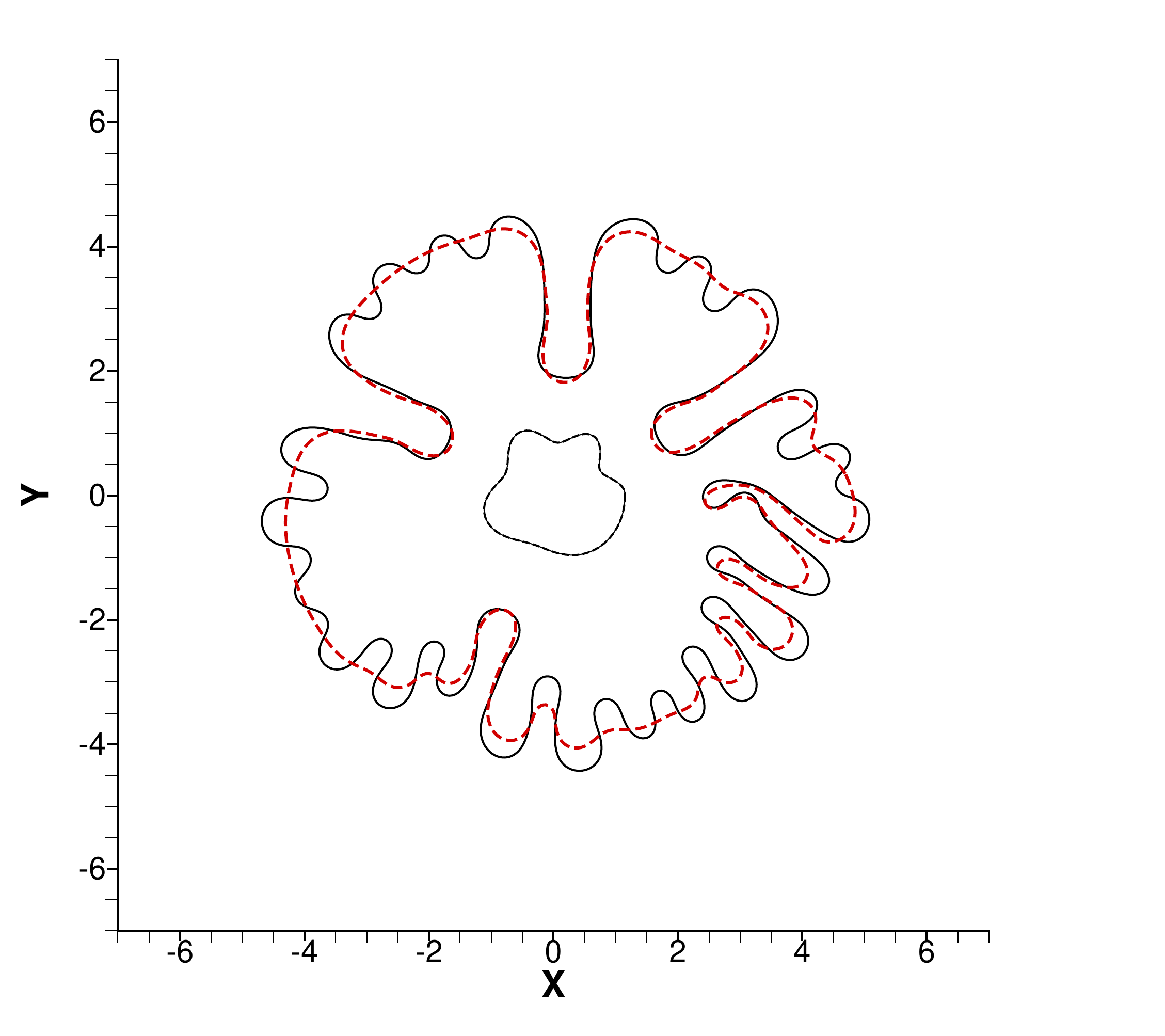}
                \caption{}
        \end{subfigure}%
        
                 \begin{subfigure}{0.4\textwidth}
                \includegraphics[width=\textwidth]{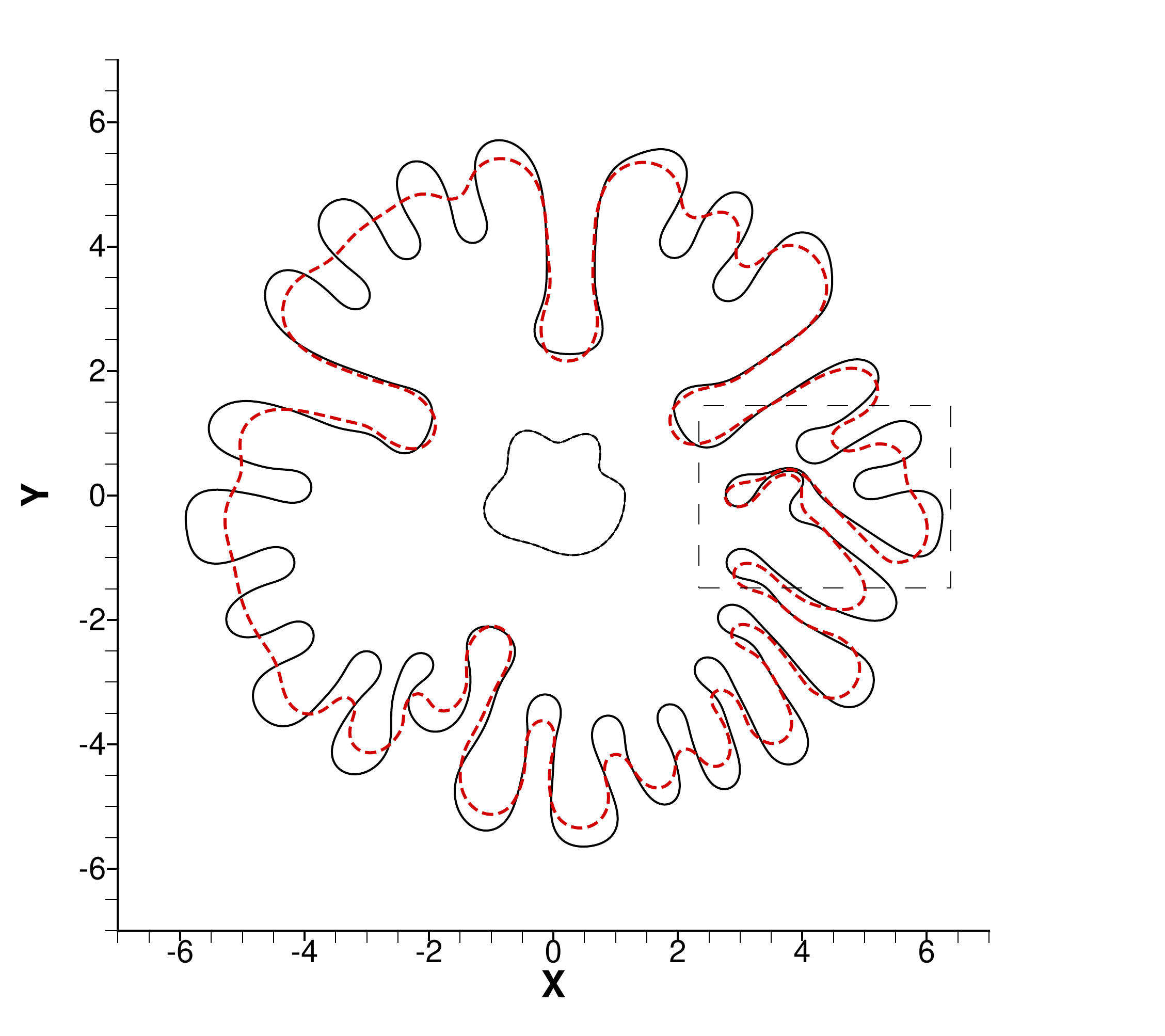}
                \caption{}
        \end{subfigure}%
        
        \caption{(Color online) Asymmetric interfacial displacement with a mobility ratio of 10. (a) t = 20, (b) t = 45, (c) t = 70. Dashed box represents zoomed in area for Fig (\ref{zoomed_mob_10}). \textbf{\textemdash} $J_0 = 0$, {\color{red} \textbf{-}\textbf{-}\textbf{-}\textbf{-}} $J_0 = 3.8$}
        \label{asym_10}
\end{figure}

Time plots of the injection can be seen in Fig. \ref{asym_10}. The onset of fingering has been delayed significantly in the dynamic wetting case, especially on the more stable fingers in the left of the domain. At t = 20, the interfaces show considerable likeness, with only minor deviation from each other. However, as time progresses, the effects of the dynamic wetting layer can be seen, whereby the bifurcation of fingers are inhibited significantly. In the areas where small fingers have formed with a large curvature, dynamic wetting has only slightly hindered the growth and subsequent bifurcation (mainly in the right of the domain). The large curvature means that the effects of dynamic wetting are proportionally less pronounced compared to areas with a small curvature. 

Due to the very flat interface at the bottom right of the domain and the relatively large global capillary number, many small fingers are able to form at the first bifurcation. Dynamic wetting has delayed the fingers from forming, but the bifurcation mode is very similar. The bases of the newly generated fingers have been pushed further into the domain due to the dynamic wetting layer.

\begin{figure}
   \hspace{-30pt}
    \centering
     \includegraphics[scale = 0.35]{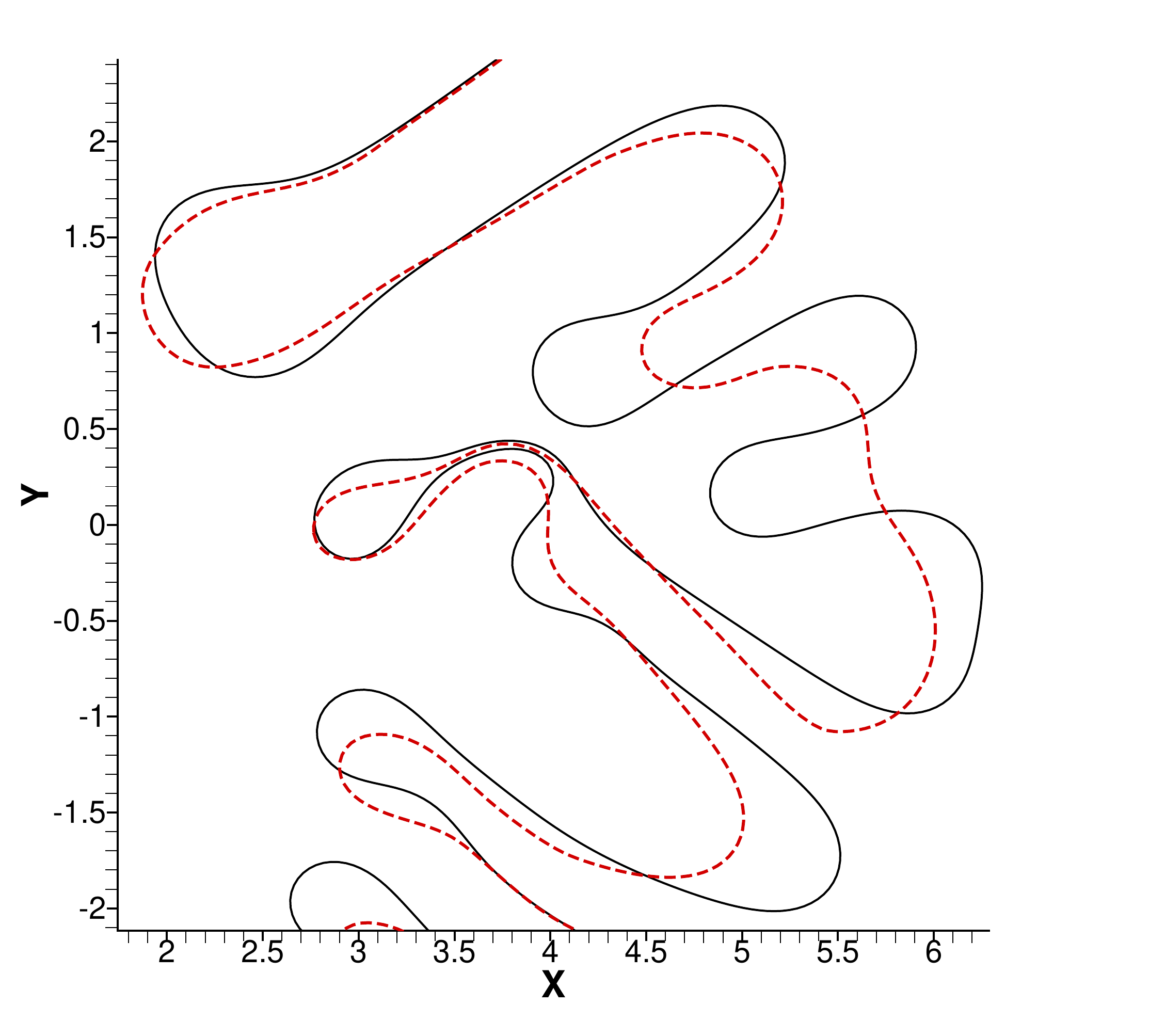}
        \caption{(Color online) Zoomed in plot of finger interaction in Fig. \ref{asym_10}. \textbf{\textemdash} $J_0 = 0$, {\color{red} \textbf{-}\textbf{-}\textbf{-}\textbf{-}} $J_0 = 3.8$}
        \label{zoomed_mob_10}
        \vspace{-10pt}
    \end{figure}

In Fig. \ref{zoomed_mob_10} we see the zoomed in area from the dashed box in Fig. \ref{asym_10}. This plot highlights the finger interaction occurring between several small fingers with large curvature and rapid growth rates. The solid line shows that without dynamic wetting, the side branching finger is thinning the base of the other primary finger significantly. This will eventually lead to finger breaking, shown in \cite{jackson2015} due to the small immiscible lubrication layer separating the fingers and the significant velocity possessed by the side branching finger. The base thinning is only slightly inhibited when dynamic wetting effects have been included (dashed line). The side branching finger is less defined than the case without dynamic wetting, but it is still advancing at nearly the same rate, and thinning the base of the other primary finger significantly. Both cases lead to finger break-off, with dynamic wetting effects delaying the base thinning mechanism but still having the same overall outcome once time has progressed. 
    \begin{figure}
\centering
           \begin{subfigure}{0.4\textwidth}
                \includegraphics[width=\textwidth]{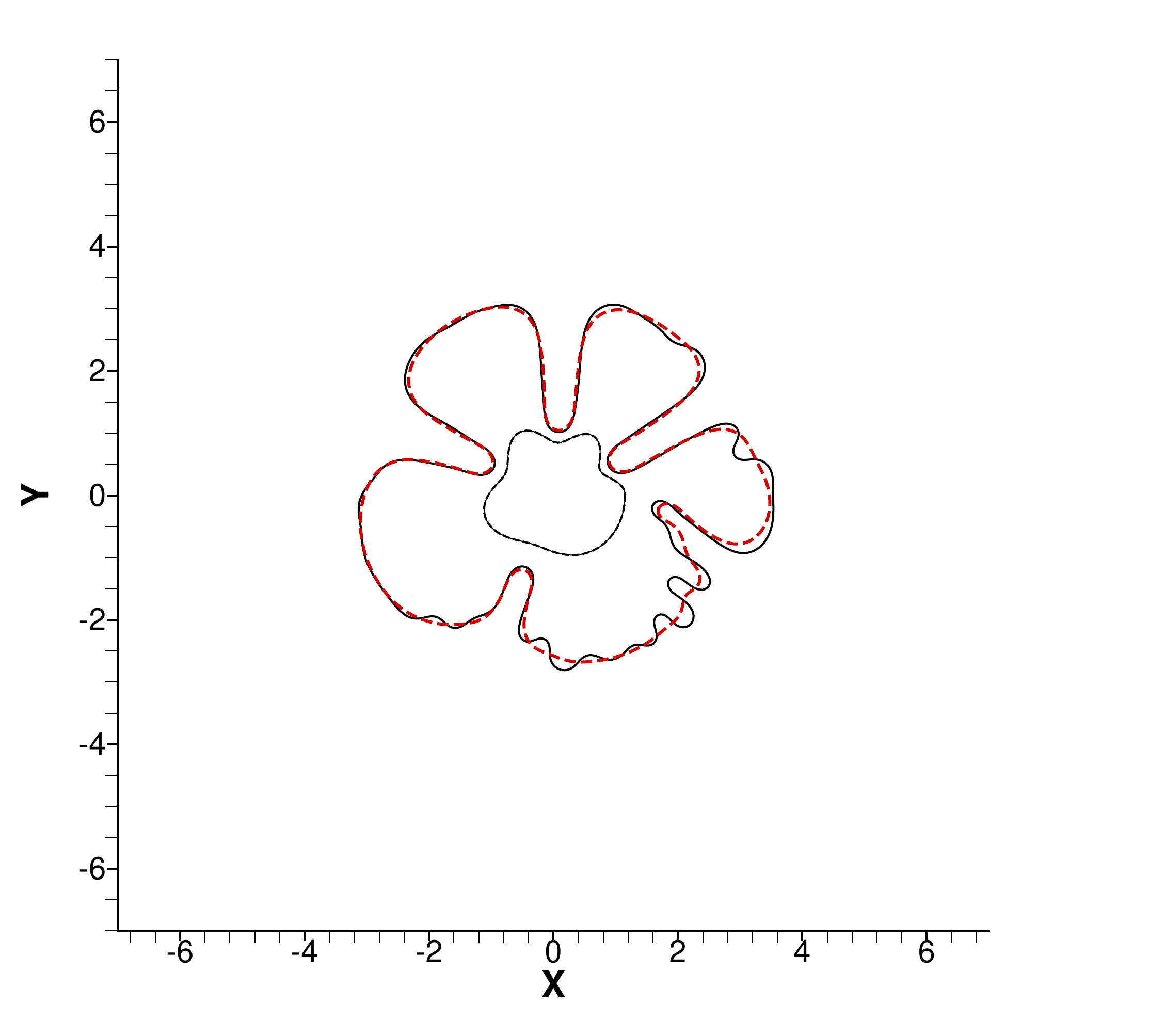}
                \caption{}
        \end{subfigure}%

         \begin{subfigure}{0.4\textwidth}
                \includegraphics[width=\textwidth]{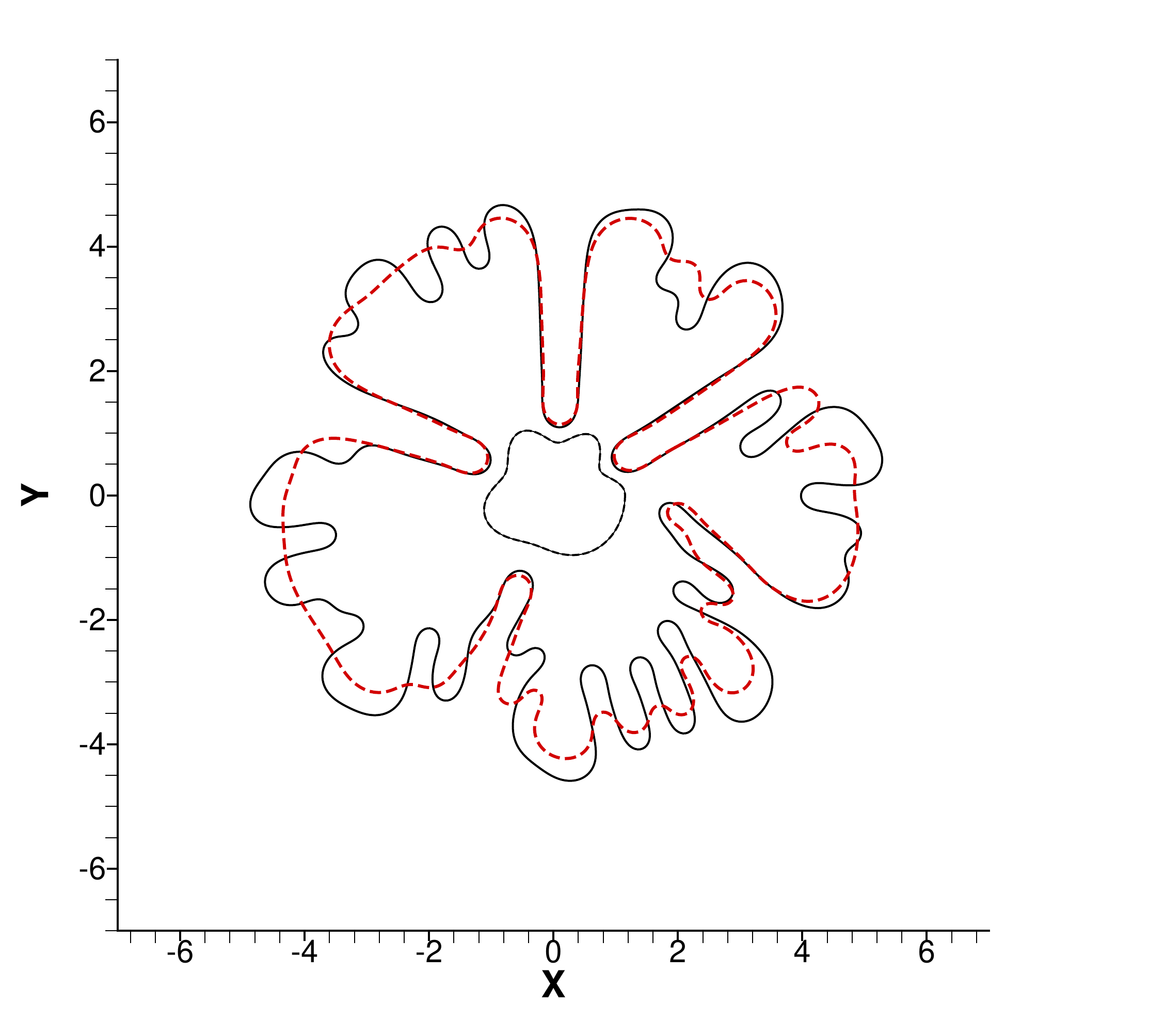}
                \caption{}
        \end{subfigure}%
        
                 \begin{subfigure}{0.4\textwidth}
                \includegraphics[width=\textwidth]{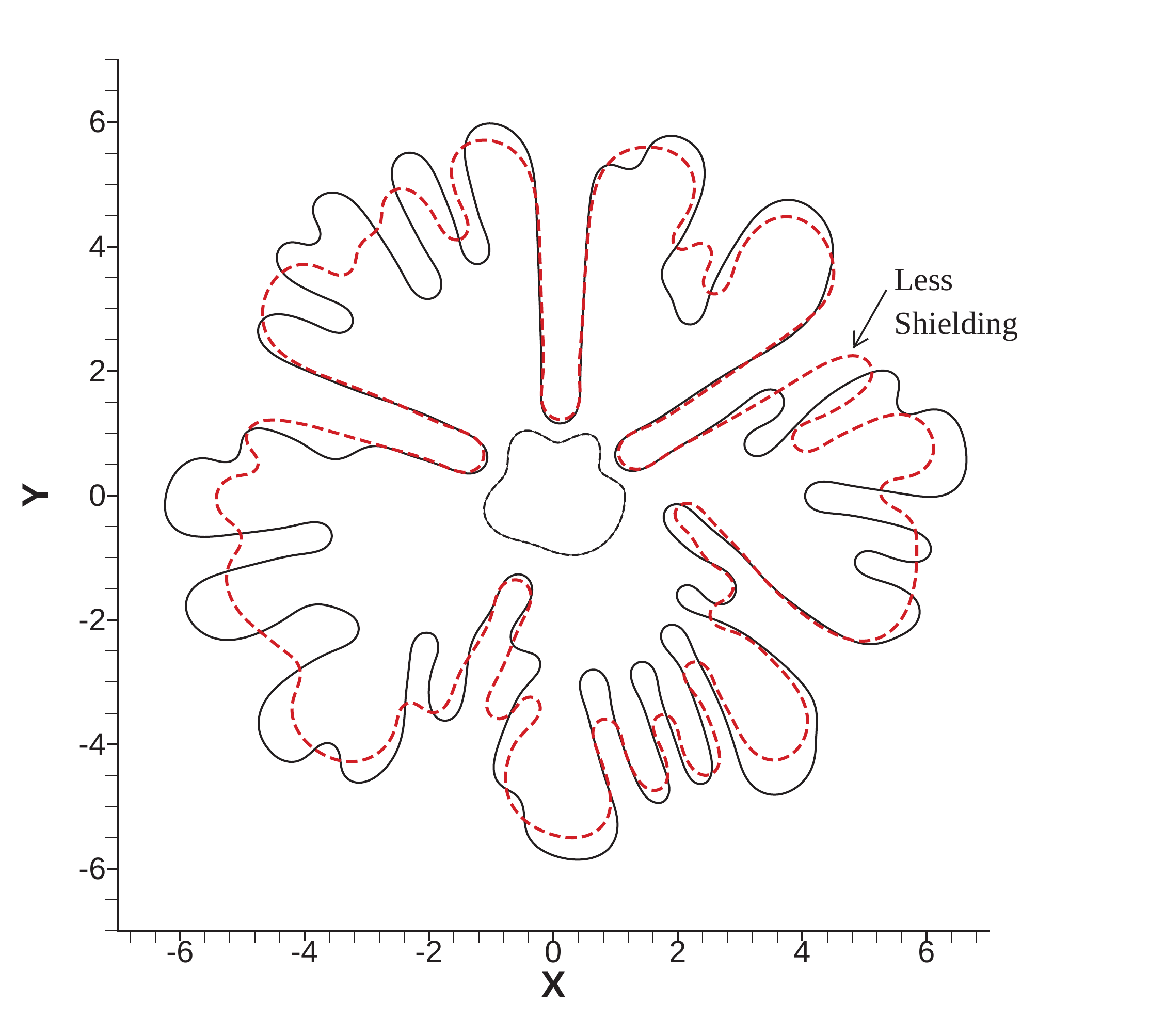}
                \caption{}
                \label{asym_c}
        \end{subfigure}%
        
        \caption{(Color online) Asymmetric interfacial displacement with a mobility ratio of 250. (a) t = 20, (b) t = 45, (c) t = 70. \textbf{\textemdash} $J_0 = 0$, {\color{red} \textbf{-}\textbf{-}\textbf{-}\textbf{-}} $J_0 = 3.8$}
        \label{asym_250}
\end{figure}
To compare the above case with a classical negligible viscosity injection, we run the same simulation but increase the mobility ratio between the two fluids to 250. This case is presented purely for comparison purposes, as most previous work in the field (when wetting effects are not considered) show flow regimes where the inner fluid has negligible viscosity. Without dynamic wetting effects, there are many previous models exploring 'single-phase' injection, representing problems such as enhanced oil recovery with air displacing oil \cite{sun2008, degregoria1986, li2007, hadavinia1995}. 

Comparing the non-dynamic wetting cases in Fig. \ref{asym_10} and Fig. \ref{asym_250} we can see the effect of increasing the mobility ratio between the fluids. Classical effects such as finger shielding are present in the high mobility ratio case, in which the secondary and side branching fingers' growth has been inhibited by the growth of primary fingers.

The combination of finger shielding and dynamic wetting have a noticeable effect on the number of fingers that are able to form in Fig. \ref{asym_250}(c). There are several fingers (notably those appearing after bifurcation on the far right primary finger) that advance more significantly into the domain than in the non-dynamic wetting case. Dynamic wetting hinders the bifurcation of the finger, which means that the initial growth of the primary finger to the side is unable to shield it's growth (shown by the note in Fig. \ref{asym_250}(c)). This means that later on in time when the finger does eventually bifurcate, the primary finger has already grown beyond it, and there is much less competition. The smaller finger can then grow relatively unhindered, and become a primary finger advancing at the forefront of the plume. 

Raising the mobility ratio has introduced the shielding effect to the viscous fingering problem, but this has been somewhat inhibited by the inclusion of the dynamic wetting layer. The delaying of bifurcation means that several previously competing fingers are not present until later in the simulation, when the primary fingers have advanced much further into the domain. The primary fingers do not compete as directly with these newly formed fingers, as the fronts are further into the domain and not advancing at the same rate as the new fingers. 

Finger competition was also found to be reduced by Dias and Miranda in the weakly non-linear stages of finger growth \cite{dias2013}. They measure finger variability in the domain and find that dynamic wetting produces fingers with a more uniform radial extent, i.e., competition has been reduced. We observe similar features here in the non-linear stages of finger growth, whereby the shielding effect is reduced due to delayed bifurcation. Fingers compete less with each other and the difference in radial extent of the different finger fronts is reduced. 

Fig. \ref{asym_10} and \ref{asym_250} show considerable likeness to the experimental results presented in Fig. 2 of \cite{ward2011}. Ward and White analyse the trailing film thickness left by a glycerol-water mix displaced by air, studying the bursting time, average film thickness and total area. Although finger morphology is not the focus of the paper, their qualitative results show remarkably similar fingering patterns to those found here, especially at the high mobility ratios presented in Fig. \ref{asym_250}. 

\section{Conclusion}

A direct boundary element method and Picard iteration scheme have been developed for the solution of immiscible displacement in a Hele-Shaw cell with non-linear wetting effects. The numerical method allows accurate and efficient evaluation of the interface velocity, permitting long term exploration of the effects of dynamic wetting in the non-linear regime of viscous finger bifurcation and competition. 

We find that during the non-linear stages of interfacial evolution, dynamic wetting delays the onset of finger splitting and changes the fundamental mode of bifurcation. Dynamic wetting smooths the interface, creating a less convoluted surface that takes longer to bifurcate. The smoothing of the interface also means that, once bifurcation starts to occur, the number of fingers produced is generally different to that when no dynamic wetting is included. 

Asymmetric interface simulations reveal that finger interaction is reduced by the inclusion of dynamic wetting in low mobility regimes and that previously found finger breaking mechanisms can be delayed but can never be completely inhibited. The dynamic wetting layer thickness decreases with time, meaning that eventually classical viscous fingering patterns emerge in the solution.    

When the mobility ratio is raised to simulate negligible viscosity fluid injection we find that finger shielding is less prominent. Secondary and side branching fingers appear later in the simulation due to delayed bifurcation meaning less competition is encountered with the primary growing fingers.

\section{Acknowledgements}

The authors would like to thank the University of Nottingham HPC team for the use of the Minerva supercomputer cluster. The present work has been partially supported by an EPSRC post-graduate research scholarship and the European Commission project PANACEA (Project Reference 282900), seventh framework program.


\begin{thebibliography}{10}

\bibitem{saffman1958}
P.G. Saffman and G.I. Taylor.
\newblock {\em Proc. R. Soc. London Ser. A}, \textbf{245}, (1958).

\bibitem{lovoll2005}
G.~Lovoll, Y.~Meheust, K.J. Maloy, E.~Aker, and J.~Schmittbuhl.
\newblock {\em Energy}, \textbf{30}, (2005).

\bibitem{lenormand1988}
R.~Lenormand, E.~Touboul, and C.~Zarcone.
\newblock {\em J. Fluid Mech.}, \textbf{189}, (1988).

\bibitem{cavanagh2014}
A.J. Cavanagh and R.S. Haszeldine.
\newblock {\em Int. J. Gre. Gas Cont.}, \textbf{21}, (2014).

\bibitem{weitz1987}
D.A. Weitz, J.P. Stokes, R.C. Ball, and A.P. Kushnick.
\newblock {\em Phys. Rev. Lett.}, \textbf{59}(26), (1987).

\bibitem{Dong2010a}
B.~Dong, Y.Y. Yang, W.Z. Li, and Y.C. Song.
\newblock {\em J. Bion. Eng.}, \textbf{7}(3), (2010).

\bibitem{stokes1986}
J.P. Stokes, D.A. Weitz, and J.P.
\newblock {\em Phys. Rev. Lett.}, \textbf{57}(14), (1986).

\bibitem{park1984}
C.W. Park and G.M. Homsy.
\newblock {\em J. Fluid Mech.}, \textbf{139}, (1984).

\bibitem{fairbrother}
F.~Fairbrother and A.E. Stubbs.
\newblock {\em J. Chem. Soc}, \textbf{1}, (1935).

\bibitem{bretherton}
F.P. Bretherton.
\newblock {\em J. Fluid Mech.}, \textbf{10}(166), (1961).

\bibitem{ratulowski}
J.~Ratulowski and H.-C. Chang.
\newblock {\em J. Fluid Mech.}, \textbf{210}, (1990).

\bibitem{daripa2011}
P.~Daripa and G.~Pasa.
\newblock {\em J. Stat. Mech.}, (L02003), (2011).

\bibitem{schwartz1986}
L.~Schwartz.
\newblock {\em Phys. Fluids}, \textbf{29}(3086), (1986).

\bibitem{reinelt1987}
D.A. Reinelt.
\newblock {\em Phys. Fluids}, \textbf{30}(2617), (1987).

\bibitem{maxworthy1989}
T.~Maxworthy.
\newblock {\em Phys. Rev. A}, \textbf{39}(11), (1989).

\bibitem{martyushev2011}
L.M. Martyushev and A.I. Birzina.
\newblock {\em J. Phys.: Condens. Matter}, \textbf{20}(045201), (2008).

\bibitem{anjos2013}
P.H. Anjos and J.A. Miranda.
\newblock {\em Phys. Rev. E}, \textbf{88}(053003), (2013).

\bibitem{dias2013}
E.~O. Dias and J.A. Miranda.
\newblock {\em Phys. Rev. E}, \textbf{88}(013016), (2013).

\bibitem{patterson1981}
L.~Patterson.
\newblock {Radial fingering in a Hele Shaw cell}.
\newblock {\em Journal of Fluid Mechanics}, 113:513 -- 529, 1981.

\bibitem{alvarez2006}
E.~Alvarez-Lacelle, J.~Ortin, and J.~Casademunt.
\newblock {\em Phys. Rev. E}, \textbf{74}(025302(R)), (2006).

\bibitem{carillo1999}
L.~Carillo, J.~Soriano, and J.~Or\'{t}in.
\newblock {\em Phys. Fluids}, \textbf{11}(778), (1999).

\bibitem{anjos2014}
P.H. Anjos and J.A. Miranda.
\newblock {\em Soft Matt.}, \textbf{10}(7459), (2014).

\bibitem{jackson2015}
S.J. Jackson, D.~Stevens, H.~Power, and D.~Giddings.
\newblock {\em Int. J. Num. Meth. Fluids}, to be published, (2015).

\bibitem{homsy1987}
G.M. Homsy.
\newblock {\em Annu. Rev. Fluid Mec.}, \textbf{19}, (1987).

\bibitem{jawson1977}
M.A. Jaswon and G.T. Symm.
\newblock {\em Integral Equations Methods in Potential Theory and
  Elastostatics}.
\newblock Academic Press, New York, (1977).

\bibitem{power1995}
H.~Power and L.C. Wrobel.
\newblock {\em Boundary Integral Methods in Fluid Mechanics}.
\newblock Computational Mechanics Publications, Southampton, (1995).

\bibitem{cabral1990}
J.J.S.P Cabral, L.C. Wrobel, and C.A. Brebbia.
\newblock {\em Eng. Ana. Bound. Ele.}, \textbf{7}(3), (1990).

\bibitem{bear1972}
Jacob Bear.
\newblock {\em Dynamics of Fluids in Porous Media}.
\newblock Dover, (1972).

\bibitem{sun2008}
Y.~Sun and C.~Beckermann.
\newblock {\em Physica D}, \textbf{237}(23), (2008).

\bibitem{degregoria1986}
A.J. DeGregoria and L.W. Shwartz.
\newblock {\em J. Fluid Mech.}, \textbf{164}, (1986).

\bibitem{li2007}
S.~Li, J.S. Lowengrub, and P.H. Leo.
\newblock {\em J. Comp. Phys.}, \textbf{225}(1), (2007).

\bibitem{hadavinia1995}
H.~Hadavinia, S.G. Advani, and R.T. Fenner.
\newblock {\em Eng. Anal. Bound. Ele.}, \textbf{16}(2), (1995).

\bibitem{ward2011}
T.~Ward and A.R. White.
\newblock {\em Phys. Rev. E}, \textbf{83}(046316), (2011).

\end{thebibliography}
\end{document}